\documentclass[aps,prl,twocolumn,superscriptaddress]{revtex4-1}
\usepackage{amsmath}
\usepackage{color}
\usepackage{graphicx}
\usepackage{epstopdf}
\usepackage{color}
\begin{document}

\title{Measurement of structured tightly focused vector beams with classical interferometry}%}%

\author{Isael A. Herrera-Hern\'andez}
\author{Pedro A. Quinto-Su}%
 \email{pedro.quinto@nucleares.unam.mx}
\affiliation{%
 Instituto de Ciencias Nucleares, Universidad Nacional Aut\'onoma de M\'exico, Apartado Postal 70-543, 04510, Cd. Mx., M\'exico.
}

%% To be edited by editor
% \dates{Compiled \today}
%\ociscodes{(140.3490) Lasers, distributed feedback; (060.2420) Fibers, polarization-maintaining;(060.3735) Fiber Bragg gratings.}
%\ociscodes{(350.4855)  Optical tweezers or optical manipulation; (140.7010)   Laser trapping; (090.1995)   Digital holography;  (060.4250)   Networks. } 
%% To be edited by editor
% \doi{\url{http://dx.doi.org/10.1364/XX.XX.XXXXXX}}
%\maketitle
%\linenumbers

\begin{abstract}
We report the first measurement with no approximations of the full field of tightly focused vector beams (NA up to $1.23$) across areas of $\sim 9-49\,\lambda ^2$. 
Prior to focusing, the structured laser light has a linear or circular polarization state.
The transverse components are extracted directly from 12 interferograms using 4 step interferometry, while the longitudinal component is extracted from the transverse fields with Gauss law. 
Different structured beams are measured to demonstrate that the method works for any field geometry. In the case of circular polarization we use vortex beams to verify the appearance of spin-orbit coupling in the axial component. 
The measurements are compared with simulations with normalized cross correlations that yield mean values $\geq 0.8$, confirming good agreement. 

\end{abstract}
%%%%%%%%%%%%%%%%%%%%%%%%%%%%%%%%%%%%%%%%%%%
\maketitle

Many important optical tools to control and interact with matter rely on pronounced focusing of laser light. Upon focusing, the fields acquire measurable 3D polarization components with sub-wavelength structures in amplitude and phase.
The first measurements of tightly focused optical fields were done with  fluorescent molecules \cite{R1d7} to probe the longitudinal polarization components.
Recently, similar approaches attempting single shot measurements \cite{R1d4} have used the properties of 4D materials, where the fluorescence emission depends on the polarization state, amplitude and phase of the light fields. However, with this approach it is only possible to extract qualitative information about the combined amplitude of the 3D polarization components. 

Currently, the state-of-the-art methods to measure highly focused vector beams rely on scanning nanoprobes like nanometer optical fiber tips in near field optical microscopy \cite{R1d2, R1d21} or probing the field with a single nanoparticle attached to a substrate \cite{R1d5, R1d6, R1d41}. However, these methods have large limitations in the quality of the reconstructed fields and in the size of the scanned areas (few wavelengths squared $\sim 4\lambda ^2$). In particular, the scanning nanoparticle approximates the field with multipole expansions, making it dependent on the field geometry, the order of the expansion and the step size of the scan ($\sim \lambda /20$) \cite{tesis1}. As a result, it is difficult to get reasonable field reconstructions.
Furthermore, many applications are based on beams with arbitrary structures with incredibly rich sub-wavelength phase patterns that span areas of tens of $\lambda ^2$. 

In the last few decades of attempts to measure these highly focused optical fields, metrology tools that are considered `paraxial' (like classical interference) have been disregarded \cite{halina2017, OtteRev, R1d4} due to the nano-scale structure and 3D polarization of tightly focused vector beams. 

Here we solve for the first time the measurement problem of tightly focused vector fields with no approximations. Our results show that complete, simple and robust measurements of these beams are finally achievable with classical interference. So far, classical interference is the only tool that can measure these fields with no approximations.

To calculate the field of a tightly focused beam in the neighborhood of the geometrical focus we use the Wolf-Richards integral \cite{R1} as a Fourier transform to speed the calculations \cite{fastfocal}. For an incident beam polarized in $x$ direction (horizontal) we have in cartesian coordinates 
\begin{equation}
\textbf{E}(x,y,z)=\frac{ife^{-ikf}}{2\pi} \textbf{IFT}[\Theta(k_x,k_y) \textbf{E}_{\infty}^{x} \frac{1}{k_z}e^{i k_z z}]
\label{eqn1}
\end{equation}
where \textbf{IFT}[ ] is the inverse Fourier transform, $k=n_2 k_0$, $k_0=2\pi/\lambda$ , $n_2=1.518$ (immersion oil) the refractive index after the lens, $k_z=\sqrt{k^2-k_x^2-k_y^2}$, $k_x$, $k_y$ are the spatial frequencies, $f$ is the effective focal length,  $\Theta(k_x,k_y)$ is the pupil function that filters spatial frequencies greater than $k_0 NA$ and $\textbf{E}_{\infty}^{x}$ is given by \cite{novotnybook}
\begin{equation}
\textbf{E}_{\infty}^x=\sqrt{\frac{n_1 k_z}{n_2 k}} \frac{E_{inc}^x\left(k_x,k_y\right)}{k_x^2+k_y^2}\cdot  
\begin{bmatrix}
           k_y^2+k_x^2 k_z/k\\
            -k_x k_y +k_x k_y k_z /k\\
           -(k_x^2+k_y^2)k_x/k
\end{bmatrix} 
\end{equation}
where $E_{inc}=A(k_x,k_y)e^{i\phi(k_x,k_y)}$, $A(k_x,k_y)$ is the incoming amplitude and $\phi(k_x,k_y)$ is the phase structure (imprinted by a spatial light modulator in the experiment),  $n_1=1.0$  (air) is the refractive index of the media before the lens (simulation details in Sec. S1 of the Supplemental Material \cite{supplementalmaterial}).

\begin{figure}
\includegraphics[width=3.5in]{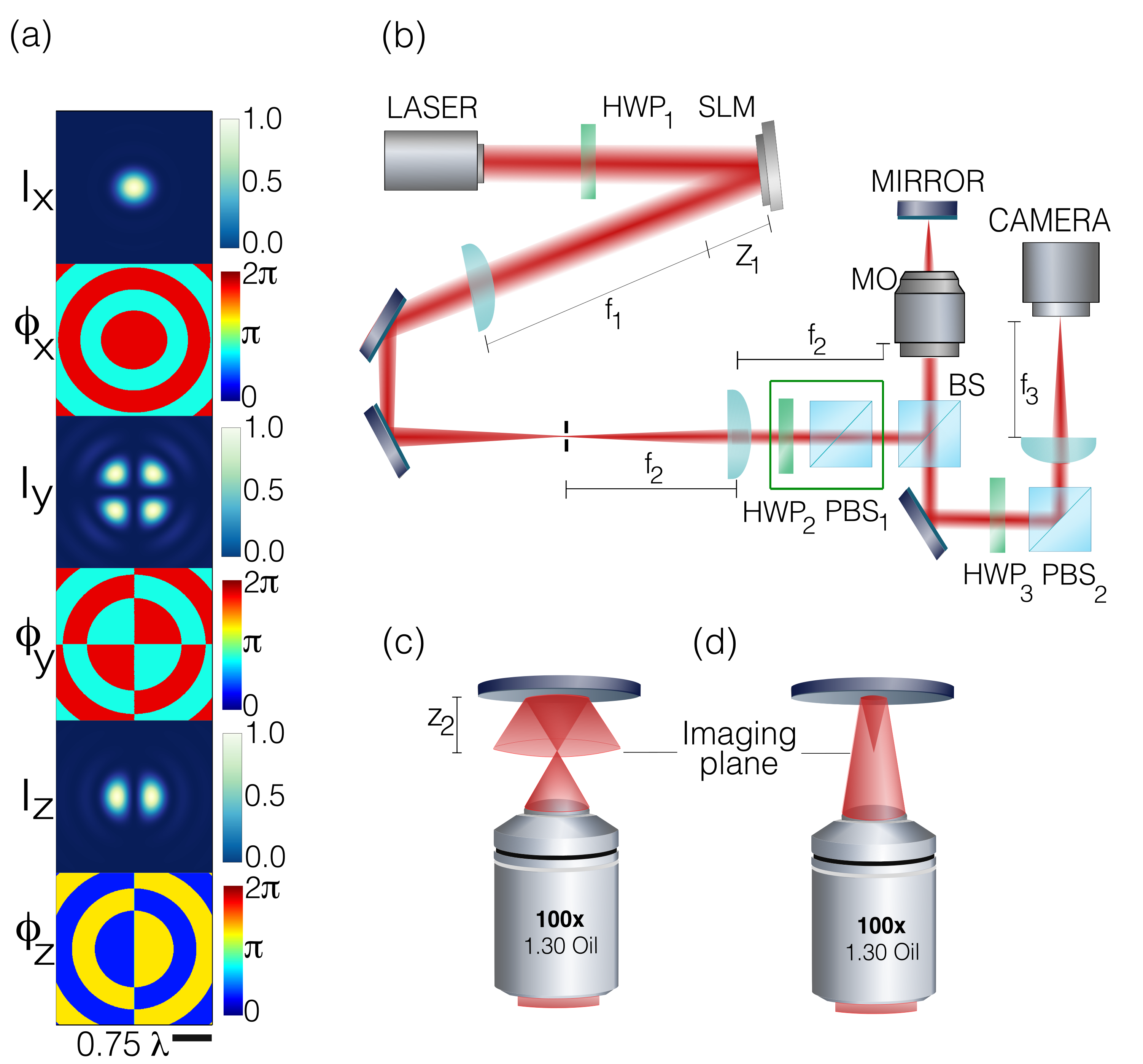}
\caption{ (a) Intensity $I_j$ and phase $\phi _j$ ($j=x,y,z$) structure of a simulated focused Gaussian beam with incident horizontal polarization calculated at $z=0\,\mu$m (focal plane). 
(b) Experimental setup: Half wave plates (HWP) and polarizing beam splitter cubes (PBS). Phase-only spatial light modulator (SLM) that modulates the phase of horizontally polarized light. Beam reducer $f_1=40\,$cm, $f_2=20\,$cm and $z_1=4\,$cm. HWP$_2$ and PBS$_1$ mounted on a retractable mount. 
(c), Reference beam reflected by a mirror displaced by $z_2$ above the imaging plane. (d), Structured beam reflected by the mirror and displaced axially  by $\Delta z=2z_2$ so that the waist is located at the imaging plane. } 
\end{figure}

The simplest vector beam is a linearly polarized Gaussian laser beam (with constant $\phi(k_x,k_y)$) focused by a high numerical aperture (NA) microscope objective. 
Figure 1(a) depicts the resulting amplitude 
and phase of this beam with incident horizontal polarization. 
After the beam is focused, the incident horizontal polarization component (x) prevails and has about $86.3\%$ of the total power, while (y) has $0.4\%$ and the axial about $13.2\%$.  
The horizontal x-component focuses into a single Gaussian spot, the y-component concentrates at 4 spots and the axial has a dipole like structure with two maximums. The phases are concentric rings or ring segments with constant phase alternating between two values with a difference of $\pi$. 
When the Gaussian beam is polarized in the other transverse component (y), the resulting transverse fields are inverted with the Gaussian structure appearing in the incident component. 

Tightly focused linearly polarized Gaussian beams have been fundamental
for many applications like optical tweezers, but so far there are no reasonable measurements \cite{tesis1, gaussianprl}. Probably because of the extreme power differences between the polarization components and the phase discontinuities across straight lines that are not compatible with the approximations of the nanoprobe-based methods.
%%%%%%%%%%%%%%%%%%%%%%%%%%%%%%%%%%%%%%%%%%%%%
%%%%%%%%%%%%%%%%%%%%%%%%%%%%%%%%%%%%%%%%%%%%%
%%%%%%%%%%%%%%%%%%%%%%%%%%%%%%%%%%%%%%%%%%%%%%
\begin{figure}
\includegraphics[width=3.2in]{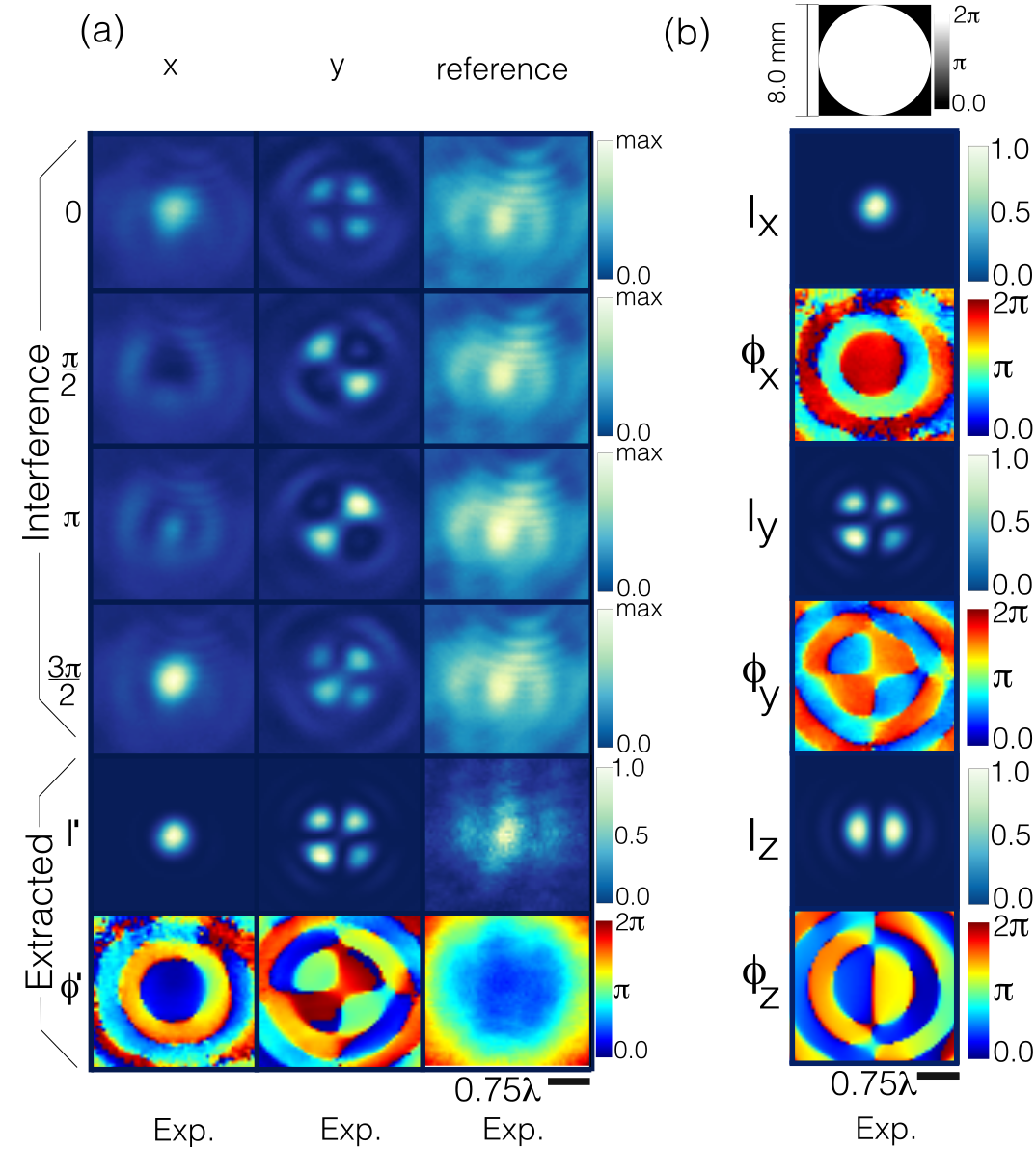} 
\caption{(a) The first 4 rows show the 12 interferograms for the x, y polarization components and the reference beam (4 each) for phase shifts of $0$, $\pi/2$, $\pi$, $3\pi /2$. Last two rows: extracted intensity and phase for the x and y components and for the reference beam $\phi '_x$, $\phi _{ref}$. (b), Reconstructed fields. The effective NA is 1.23}.
\end{figure}
%%%%%%%%%%%%%%%%%%%%%%%%
%%%%%%%%%%%%%%%%%%%%%%%%%%%%%%%%%%%%%%%%%%%%%f3

Our experiment is performed in a standard holographic optical tweezers setup (Fig. 1(b)) with a few added polarization elements (half wave plates HWP and polarizing beam splitter cubes PBS). The light source is a linearly polarized continuous laser with a wavelength of 1064\,nm (Gaussian profile) with the polarization state controlled by the first half wave plate HWP$_1$. The phase-only SLM imprints a 2D phase mask to horizontally polarized light. In this way, the light reflected by the SLM is a horizontally polarized modulated beam and a vertically polarized reference (Instrument details in Sec. S2 of the Supplemental Material \cite{supplementalmaterial} )

The size of both beams is reduced with a pair of lenses, then they can propagate through HWP$_2$ and a polarizing beam splitter cube (PBS$_1$) installed on a retractable mount. Finally, the beams are reflected by a 50/50 non polarizing beam splitter cube into the back aperture of the microscope objective (100x, 1.3 NA), where the beams have a waist of $1.86\,$mm.
After both beams are focused by the microscope objective, they are reflected back by a mirror mounted on a piezo stage that controls the distance $z_2$ to the imaging plane. The image is projected by a lens into a USB-CMOS camera and the transverse polarization state of the image is selected with HWP$_3$ and PBS$_2$ .

The reference beam focuses at the imaging plane (Fig. 1(c)) and then it is reflected back by the mirror and imaged as a diverging beam. 
The structured beam is displaced vertically along the optical axis by $\Delta z=2z_2 \sim 5-8\, \mu$m (Fig. 1(d)), so that the waist is imaged by the microscope objective. This is done by adding a displacement phase to the digital hologram $\phi _d (k_x,k_y)=-\Delta z k_z $. This phase that has been used in microscopy \cite{fasedes, fasedesnat} and translates the beam by $\Delta z$ without adding spherical aberrations (see Sec. S2 of the Supplemental Material \cite{supplementalmaterial}).

%%%%%%%%%%%%%%%%%%%%%%%%%%%%
\begin{figure*}
\includegraphics[width=7.0in]{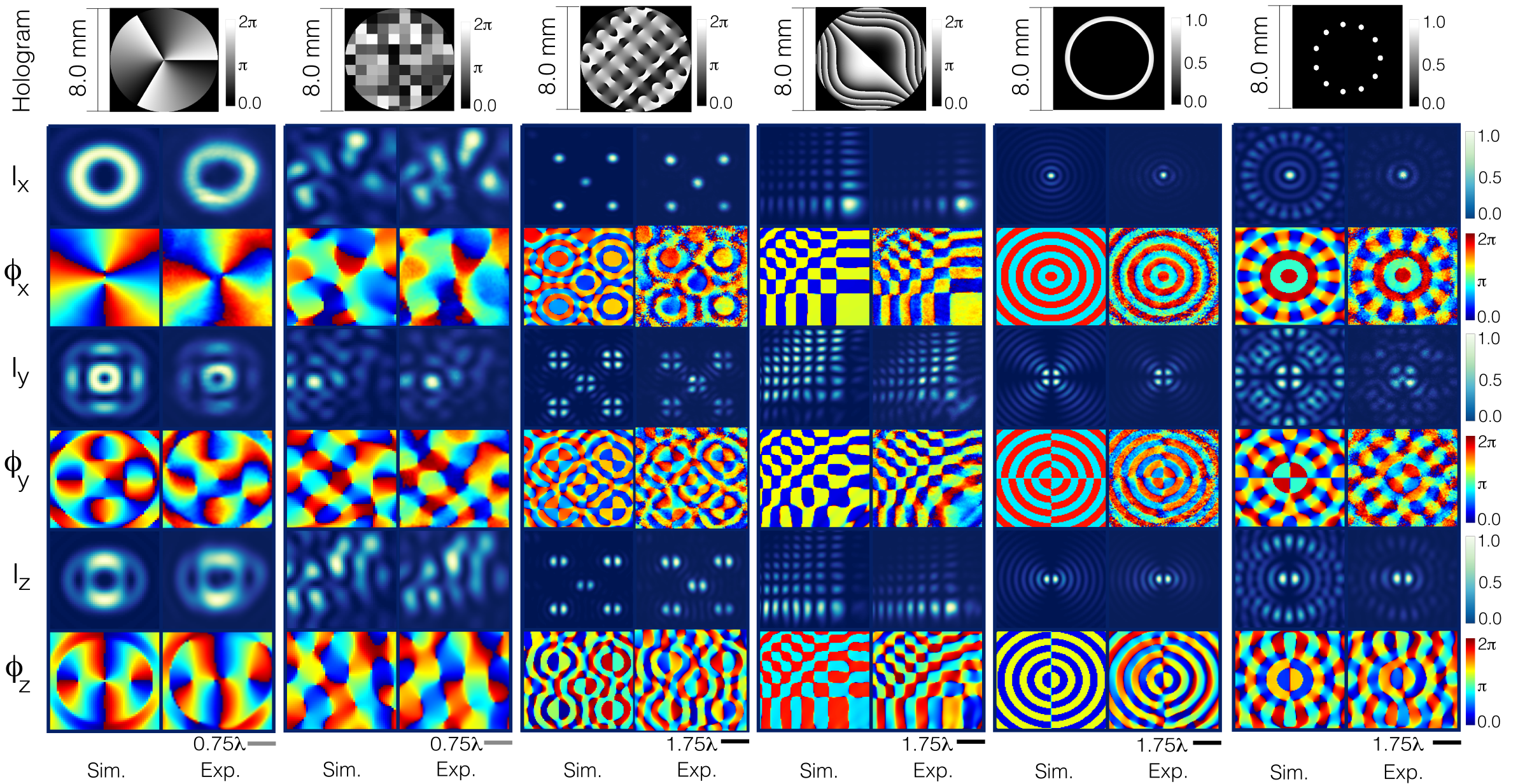} %6.8
\caption{The results are arranged in column pairs for each structured beam, the left column is the simulated vector beam and the right has the measurements. The top row depicts the holograms or the aperture at the SLM. The fields are arranged in the following order (top to bottom): $I_x$, $\phi _x$, $I_y$, $\phi _y$, $I_z$, $\phi _z$. 
The structured beams are (left to right) vortex of order 3, speckle, five focused spots generated with the GS algorithm, Airy, Bessel, and a discrete beam with 11 sources. Average power distribution for the first 3 cases is (86.3\% 0.4\%, 13.2\%), Airy: (87.1\% 0.4\%, 12.4\%). Amplitude modulated: Bessel (75.3\%, 0.9\%, 23.7\%), discrete (86.9\% 0.2\%, 12.8\%). The images have a size of $3\times 3 \lambda^2$ (first two cases) and $7\times 7\lambda^2$. All the intensities are normalized with respect to the maximum in each case. $\Delta z =5\,\mu$m with the exception of the Airy beam where $\Delta z =8\,\mu$m. The effective NAs for the different beams are: 1.23 (first four cases), 1.1 (Bessel) and 0.8 (discrete).   
 } 
\end{figure*}
%%%%%%%%%%%%%%%%%%%%%%%%%%%%%%%%%%%%%%%%%%%%%%%%%%%%%%%%%%%%%%%%
%%%%%%%%%%%%%%%%%%%%%%%%%%%%%%%%%%%%%%%%%%%%%%%%%%%%%%%%%%%%%%%%%%%%%%%%

We measure 12 interferograms: 8 for the transverse polarization components and 4 for the reference beam. The procedure is illustrated in Fig. 2 for the case of a focused Gaussian beam that has the same properties as the one simulated in Fig. 1(a) and can be compared with the results shown in \cite{tesis1, gaussianprl}. We choose the same Gaussian reference for both transverse polarization components in the following way:
for the x component, we rotate the polarization of both beams (after the SLM) with HWP$_2$ and project them into a horizontal polarization state with PBS$_1$. To measure the acquired y component of the modulated beam we remove HWP$_2$ and PBS$_1$ (the phase difference between modulated and reference beams is not affected) so that upon focusing, the structured beam acquires a vertical polarization component which can interfere with the Gaussian structure of the reference beam (incident with vertical polarization).

The interference for each transverse polarization component $j=x,y$ with the reference beam is imaged in 4 cases $\mathcal{I}_{ij}$ (four step phase shift interferometry \cite{ref4fases1, ref4fases}, see  Sec. S2 of the Supplemental Material \cite{supplementalmaterial}), where $\mathcal{I}_{ij}$ represents the interferograms in the $j$ polarization channel with an added phase shift to the incident structured beam of $\Delta \phi _i$ (i=1-4) with values of $0,\,\pi/2 ,\,\pi ,\, 3\pi /2$. These images (Fig. 2(a), first 4 rows, first 2 columns) are used to extract the product of the intensities $I' _j=I_j I_{ref}$ and the phase difference $\phi _j'=\phi _{j} -\phi _{ref}$ between reference ($I_{ref}$, $\phi _{ref}$) and structured beam ($I_j$, $\phi _j$). 
The phase difference is extracted with $\phi '_j (x,y) =tan^{-1}[(\mathcal{I}_{j4} -\mathcal{I}_{j2}) /(\mathcal{I}_{j1}-\mathcal{I}_{j3})]$, while the modulated intensity is given by $I'_{j}=((\mathcal{I}_{j1} -\mathcal{I}_{j3})^2 +(\mathcal{I}_{j2}-\mathcal{I}_{j4})^2)$ Fig. 2(a), last two rows first two columns. 

Notice that $\phi '_j$ is wrapped in the diverging phase of the Gaussian reference beam. In this way, the unwrapped phases are obtained by subtracting the phase of the reference beam $\phi _{ref}$ so that $\phi _j= mod(\phi '-\phi _{ref}, 2\pi)$. The intensities $I_j$ are obtained with $I_j=I' _j/I_{ref}$ (Fig. 2(b)).
The reference phase $\phi _{ref}$ and intensity $I_{ref}$ (Fig. 2(a) last column) are also obtained from 4 step phase shift interferograms $\mathcal{I}_{ref-i}$ with a collimated structured beam (Sec. S2 of the Supplemental Material \cite{supplementalmaterial}). %
In the experiment, we follow the procedure for the x-component (with HWP$_2$ and PBS$_1$). A lens phase is projected onto the SLM that focuses the structured beam at $z_1$, in this way the modulated beam is collimated by the microscope objective and it is possible to extract the properties of the reference. The unwrapped fields are in Fig. 2(b). 
The effect of $z_1$ is to add a negligible perturbation that does not disrupt our measurement. This is discussed in Sec. S2 of the Supplemental Material \cite{supplementalmaterial}.

The $E_j(x,y,z)$ components are constructed by using the relation $E_j=\sqrt{I_j}e^{\phi_j}$ and the axial field is extracted from the measured transverse fields utilizing Gauss law written in the angular spectrum representation $E_z(x,y,z)=\mathbf{IFT}[-\frac{1}{k_z}(k_x \hat{E}_x(k_x,k_y;z)+k_y \hat{E}_y(k_x,k_y;z))] $   
and  $\hat{E}_j$ is  given by: $\hat{E}_j(k_x,k_y;z)=\mathbf{FT}[E_j(x,y,z)]$, where $\mathbf{FT}[E_j(x,y,z)]$ is the direct Fourier transform. 

Next, in Fig. 3 we show measurements for a variety of widely used beams to show that it is possible to resolve arbitrary field geometries due to the lack of approximations. We explore a vortex beam ($m=3$), speckle fields ($20\times 20$ random phasors) \cite{micromanipulation3}, five foci calculated with the Gershberg-Saxton algorithm \cite{gs}, Airy \cite{micairy, holairy} (checkerboard-like phase structure), Bessel \cite{micbessel, micromanipulation2} and discrete beams \cite{optbook}.
Perhaps with the exception of the x component of the vortex beam, the rest of the fields in Fig. 3 are well outside the capabilities of current nanoprobe methods due to the field geometries (like the phase in Airy, GS), the areas spanned by the beams and the huge power imbalances across the components.
The last two cases (Bessel and discrete beams) are generated with very low efficiency holograms where the undiffracted zero order from the SLM dominates, showing that it is possible to extract the fields even in those conditions. 
In the measurements there are some regions where the measured phases have a difference of almost $\pm 2\pi$ compared with the simulations, shifting the colormaps from blue to red or other color pairs.

In order to quantify the similarity between measurements and simulations, we calculate normalized cross correlations (NCC) \cite{ncc}. For each beam, the NCC is calculated for the 3 components (amplitude and phase). 
The data is in Sec. S2 and Fig. S4 of the Supplemental Material.
We observe that there is good agreement for all the  cases with mean NCCs  $\geq 0.8$. The precision is $\lambda /44$ (25 repetitions) which is smaller than what is typically achieved in macroscopic or paraxial step interferometry \cite{4step_slm}. That is explained by the subwavelength structures of the fields which are very sensitive to nanometer variations in the position of the mirror (Fig. 1(c)).
More examples of tightly focused beams with other structures are in the Sec. S3 of the Supplemental Material \cite{supplementalmaterial}.

%%%%%%%%%%%%%%%%%%%%%%%%%%%%%%%%%%%%%%%%%%%%%%%
%%%%%%%%%%%%%%%%%%%%%%%%%%%%%%%%%%%%%%%%%%%%%%%%%%%%
\begin{figure}
\includegraphics[width=3.5in]{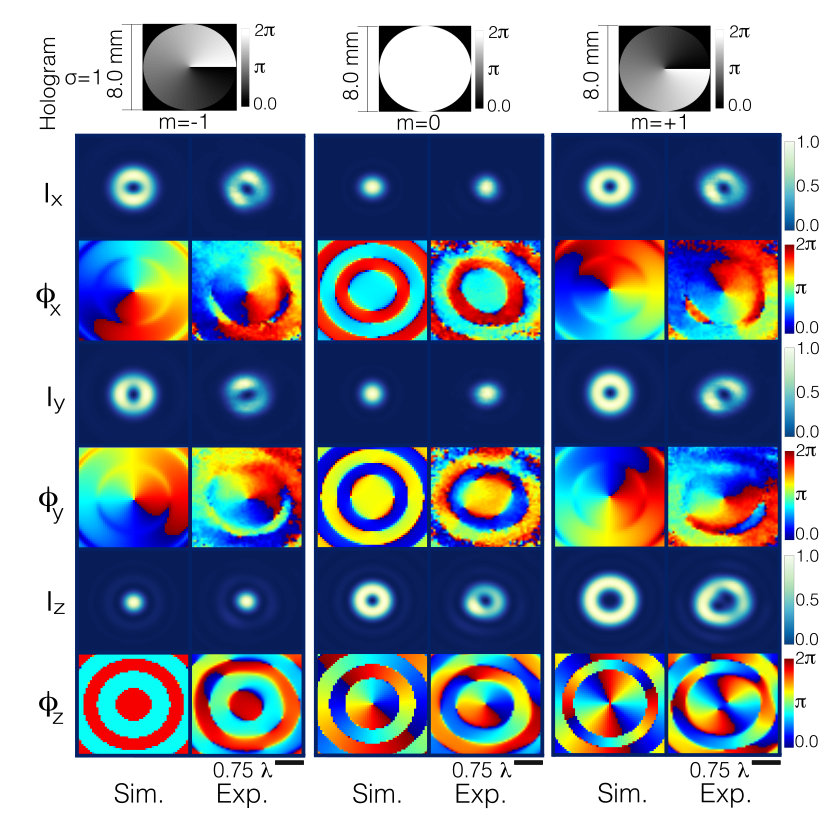} 
\caption{Circular polarization $\sigma =1$, the column pairs are $m=-1,0,+1$ (left to right). The fields are arranged in the following order (top to bottom): $I_x$, $\phi _x$, $I_y$, $\phi _y$, $I_z$, $\phi _z$. 
The images have a size of $3\times 3\,\lambda ^2$. Average power distribution per component (44.3\%, 44.5\%, 11.1\%). Effective NA of 1.23 for all cases.}
\end{figure}
%%%%%%%%%%%%%%%%%%%%%%%%%%%%%%%%%%%%%%%%%%%%%%%%%%%%%

%
Finally, we explore circular polarization states, where both transverse components have similar contributions and are dephased by $\pi/2$. In order to compare with nanoprobe measurements \cite{refr}, we chose vortex beams with different values of the charge $m$ where the phenomena of spin orbit coupling appears and the axial component acquires a charge of $m+\sigma$ \cite{refangular}.
To perform the circular polarization experiments we add a quarter wave plate after PBS$_1$ to the experimental setup.
We measure both transverse components in the same way we measure the $x$ component and we select the transverse interferograms with HWP$_3$ and PBS$_2$. The reference is extracted in the same way focusing the modulated beam at $z_1$. 
We show the experimental results  for $\sigma =+1$ in Fig. 4 (More data for  $\sigma=\pm 1$ in Sec. S3 of the Supplemental Material \cite{supplementalmaterial}). 
%%%%%%%%%%%%%%%%%%%%%%%%%%%%%%%%%%%%%%%%%%%%%%%%%%%%
%

%
To conclude, we have shown the first complete measurements of fundamental tightly focused vector beams with no approximations, resulting in unprecedented quality. Furthermore, the results were obtained with
classical interferometry, which has been disregarded as an option to measure these beams. More importantly, we have demonstrated that so far, classical interferometry is the only method that does not require approximations.
This method also should work with laser sources of different spectral widths \cite{ultrashort} and even with low visibility which is compatible with 4 step interferometry \cite{visibility}.
Another advantage of using classical interferometry is that we can measure the phase around regions of low intensity, so it will enable the experimental study of non paraxial singular fields \cite{dennisjosa}.
Our results should enable most groups that work with tightly focused beams to finally explore the full fields with small changes to their setups. 
Future work will involve extending this method to beams that have arbitrary polarization states before focusing.

%%%%%%%%%%%%%%%%%%%%%%%%%%%%%%%%%%%
\subsection*{Acknowledgements}
Work partially funded by DGAPA UNAM PAPIIT grant IN107719, CTIC-LANMAC 2020 and CONACYT LN-299057. IAHH thanks CONACYT for a scholarship.  
Thanks to Jos\'e Rangel Guti\'errez for machining some of the optomechanical mounts. 

\bibliographystyle{apsrev4-1}
%\bibliographystyle{}

%%%%%%%%%%%%%%%%%%%%%%%%%%%%%%%%%%%%%%%%%%%%%%%%%

\end{document}